\documentstyle[preprint,aps]{revtex}
\begin{document}
\title{On the WKB Quantum Equivalence between Diverse $p$--brane Actions}
\author{Stefano Ansoldi\footnote{E-mail
address: ansoldi@trieste.infn.it}}
\address{Dipartimento di Fisica Teorica,
Universit\`a di Trieste,\\
INFN, Sezione di Trieste}
\author{C.Castro\footnote{E-mail
address: castro@ctsps.cau.edu }}
\address{Center for Theoretical Studies of Physical Systems\\
Clark Atlanta University, Atlanta, GA.30314}
\author{Euro Spallucci\footnote{E-mail
address: spallucci@trieste.infn.it}}
\address{Dipartimento di Fisica Teorica,
Universit\`a di Trieste,\\
INFN, Sezione di Trieste}

\maketitle

\begin{abstract}
We consider an  action for a closed, bosonic, $p$--brane, where the brane
tension is not an assigned  parameter but rather it is induced by a maximal
rank gauge $p$--form.
This model is classically equivalent to the Nambu--Goto/Howe--Tucker model.
We investigate how this classical equivalence can be implemented in the path
integral framework. For this purpose we adopt a ``first order'' integration
procedure over gauge $p$--forms and a ``shortened'' Fadeev--Popov procedure.
\end{abstract}

\newpage

Diverse action functionals have been proposed to describe
dynamics of a relativistic, bosonic, $p$--brane \cite{neman}.
The first brane action, proposed in 1975, was a generalization of the
Nambu--Goto action for strings, i.e. the measure of the brane world
history \cite{ct}
\begin{equation}
S_{DNG}[\, Y\,] =
-m_{p+1}\int_\Sigma d^{\, p+1}\, \sigma\sqrt{ -\gamma} \ ,\quad
\gamma\equiv det\left(\,\partial_m Y^\mu \partial_n Y_\mu\,\right)
\quad ,
\label{ng}
\end{equation}
where $m_{p+1}$ is the ``$p$--tension''. We use ``$p$'' for the
spatial dimensionality of the brane; thus, the coordinates $\sigma^m$,
$m=0\ , 1\ ,\dots\, p$, span the $(p+1)$--dimensional world manifold
$\Sigma$.
The $D$ functions $Y^\mu(\sigma)$, $\mu=0\ , 1\ ,\dots\, D$, are the
brane coordinates in the $D$--dimensional target spacetime. The special
case $p=2$ and $D=4$  was already introduced in 1962 by Dirac in an
attempt to resolve the electron--muon puzzle in terms of a relativistic
membrane \cite{dng}.\\
An alternative description, preserving world manifold reparametrization
invariance, can be achieved by introducing an auxiliary world manifold
metric $g_{\, m\, n}(\sigma)$ and a ``cosmological term'' \cite{ht},
\cite{netra},
\begin{equation}
S_{HTP}[\, Y\ , g\,]= -{m_{p+1}\over 2}\int_\Sigma
d^{\, p+1}\sigma
\sqrt{ -g} \,
\left[\,g^{\, m\, n}\partial_m\, Y^\mu\, \partial_n\, Y_\mu -(p-1)\,
\right]
\quad ,
\label{spolya}
\end{equation}
where $g\equiv \det( g_{\, m\, n} )$. In both functionals (\ref{ng})
and (\ref{spolya}) the brane tension $m_{p+1}$ is a {\it pre--assigned }
parameter.\\
The two actions (\ref{ng}) and (\ref{spolya}) are classically
equivalent as the ``field equations'' ${\delta S/\delta g^{\, m\,
n}(\sigma)}=0$ require the auxiliary world metric to match the
induced metric, i.e. $g_{\, m\, n}=\gamma_{\, m\, n}=\partial_m
Y^{\,\mu} \,\partial_n Y_{\,\mu}$. Moreover they are also complementary:
$S_{DNG}$ provides an  ``extrinsic'' geometrical description in terms
of the embedding functions $Y^\mu(\sigma)$ and the induced metric
$\gamma_{mn}$,
while $S_{HTP}$ assigns an ``intrinsic'' geometry to the world
manifold $\Sigma$ in terms of the metric $g_{\, m\, n}$ and the
``cosmological constant'' $m_{p+1}$; the $Y^\mu(\sigma)$ functions enter
as a ``multiplet of scalar fields'' propagating on a curved
$(p+1)$--dimensional manifold.\\
More recently new action functionals have been proposed
where the brane tension, or world manifold cosmological constant,
is not an {\it a priori } assigned parameter, but follows from the
dynamics of the object itself and can attain both positive and vanishing
values. Either Kaluza--Klein type mechanism \cite{blt} and modified
integration measure \cite{eduard} have been proposed as candidate
dynamical processes to produce tension at the classical level.

The main purpose of this note is to investigate how the  dynamical
generation of the brane tension and the equivalence between diverse
action functionals can be extended at the quantum
level in the WKB approximation of a ``sum over histories'' approach.
However, before considering the path--integral it is instrumental
to review how classical dynamics leads to the action (\ref{ng}) as
an {\it effective,} on--shell action.

The guiding principle to assign  the $p$--brane tension the role of a
dynamical variable is borrowed from modern cosmology, where
the cosmological constant can be represented by a {\it maximal rank}
gauge  $p$--form \cite{acl}.
Thus, we introduce the following action functional
\begin{eqnarray}
S[\, Y\ , g\ , A\,] &=& -\int_\Sigma d^{p+1}\sigma\, \sqrt{- g}\left[\,
{m_{p+1}\over 2}\, {(-\gamma)\over (-g)}
 -{1\over 2(p+1)!}\, F_{m_1\dots m_{p+1}}\,
F^{m_1\dots m_{p+1}}
\,\right]\nonumber\\
&=& -\int_\Sigma d^{p+1}\sigma\, \left[\,
{m_{p+1}\over 2}\, {(-\gamma)\over \sqrt{- g}}\,
 -{\sqrt{- g}\over 2(p+1)!}
\, F_{m_1\dots m_{p+1}}\, F^{m_1\dots m_{p+1}}
\,\right]
\quad ,
\label{lschild}
\end{eqnarray}
where the world manifold $\Sigma$ has a space--like boundary
$\partial\Sigma$ whose target space image will represent
a closed, $p$--dimensional, relativistic object. Moreover, we
introduce on the world manifold  a maximal rank gauge field,
$A _{m_2\dots m_{p+1}}( \sigma )$, with
field strength $F_{m_1\dots m_{p+1}} \equiv
\partial_{[\,m_1 } A_{m_2\dots m_{p+1}\,]}( \sigma )$. To preserve
gauge invariance under $\delta A_{m_2\dots m_{p+1}}=\partial_{[\,m_2}
\Lambda_{m_3\dots m_{p+1}\,]}$ in the presence of a boundary we
must give up a current--potential interaction term and consider {\it
only} a gravitational coupling  $A$--$g$. By a suitable rescaling
of the brane coordinates the dimensional constant $m_{p+1}$ can be
washed out, and the classical action (\ref{lschild}) written without
any dimensional scale.  Our goal
is to show that the Dirac--Nambu--Goto functional can be
obtained as an {\it effective} action from (\ref{lschild}) once
the classical field equations for the $p$--form gauge potential
are solved.\\
Varying the action (\ref{lschild}) with respect to  $A$ we get
$$
{\delta S[\, Y\ , g\ , A\,]\over
\delta A_{m_2\dots m_{p+1}}(\sigma)}=0
\quad \longrightarrow \quad
\partial_m\left(\, \sqrt{- g} \,  F^{m m_2\dots m_{p+1}}\,\right)=0
\quad ,
$$
which, since $A$ is maximal on the $p$--brane, has the solution
\begin{equation}
F^{m m_2\dots m_{p+1}}= \Lambda \epsilon^{m m_2\dots m_{p+1}}=
\Lambda {1\over \sqrt{- g} }\delta^{[\, m m_2\dots m_{p+1}\,]}
\quad ,
\label{cc}
\end{equation}
where $\Lambda$ is an arbitrary integration constant. By inserting
the solution (\ref{cc}) back into (\ref{lschild}) we obtain
\begin{equation}
S[\, Y\ , g\,] = -\int_\Sigma d^{p+1}\sigma\,\left[\,
{m_{p+1}\over 2}\, {(-\gamma)\over \sqrt{-g}}
 +{\Lambda^2\over 2}\sqrt{-g}\,\right]
\quad ,
\label{son}
\end{equation}
where, the world manifold cosmological constant $\Lambda^2$
shows up as the {\it on--shell} value of the gauge field
kinetic  term.\\
The on-shell action (\ref{son}) depends from the
world metric {\it only} through the volume density
$\sqrt{-g}$. Hence, variations with respect to $g_{mn}$ reduce
to variations with respect $\sqrt{-g}$:
\begin{equation}
{\delta S[\, Y\ , g\, ]\over\delta g_{mn}(\sigma)}=0
\quad \longleftrightarrow \quad {\delta S[\, Y\ , g\, ]\over\delta
 \sqrt{-g}}=0
 \quad .
\label{tmn}
\end{equation}
By inserting the solution (\ref{cc}) into (\ref{tmn}) we get
\begin{equation}
m_{p+1}{(-\gamma)\over (-g) }
=
\Lambda^2
\quad
\Rightarrow
\quad
\sqrt{ -g}
 =
{1\over\Lambda} \sqrt{\, m_{p+1}}
\sqrt{-\gamma}
\quad
\label{gcl}
\end{equation}
and
\begin{equation}
S = - \Lambda \sqrt{\, m_{p+1} }
\int_\Sigma d^{p+1}\sigma\,\sqrt{-\gamma}\equiv -\rho_p\int_\Sigma
d^{p+1}\sigma\,\sqrt{-\gamma}
\quad .
\label{sng}
\end{equation}
After solving for the world metric  in terms of the brane coordinates,
the action (\ref{lschild}) turns out to be equivalent to a
Dirac--Nambu--Goto action with a dynamically induced brane tension
given by $\displaystyle{\rho_p\equiv \Lambda \sqrt{m_{p+1} }  }$.
Let us remark that $\Lambda$ can take any value including zero.
Accordingly, {\it null branes}, corresponding to the action
\begin{equation}
S_{null}[\, Y\ , g\,] = -{m_{p+1}\over 2} \int_\Sigma d^{p+1}\sigma
\,{(-\gamma)\over \sqrt{-g}}
\quad ,
\label{nullb}
\end{equation}
are included in our description as well.
    This special case stresses how the parameter
$m_{p+1}$ is not necessarily the brane tension, but only a dimensional
constant needed to allow the various dynamical fields in the action
to keep their canonical dimensions\footnote{ If we keep
$m_{p+1}\ne 0$ the brane coordinates
have canonical dimension of length, while the world  and the induced
metric  are dimensionless (in units $\hbar=1$, $c=1$), i.e.
$[\, Y^\mu\, ]=M^{-1}$, $[\, \gamma_{mn}\, ]=[\, g_{mn}\, ]=1$. }.

In the second part of this note we shall discuss the above equivalence
at the quantum level. The basic quantity encoding the $p$--brane
quantum dynamics is the {\it boundary wave functional,} or
vacuum---one--brane amplitude
\begin{equation}
Z\equiv Z[\widehat Y\ ,\widehat A\ , \widehat g\,] =
\int^{\widehat g }[\, Dg_{mn}\,] \int^{\widehat Y } [\, DY^\mu\,]
\int^{\widehat A }[\, DA\,]\, \exp\left(\, i S[\, Y\ , g\ , A\,]\,
\right)
\quad ,\label{zab}
\end{equation}
where the sum is over all bulk fields configurations inducing
    ``hatted'' fields on the boundary of the brane. We are assuming
    that the brane world manifold has a single, $p$--dimensional boundary,
    parametrized as $\sigma^m=\sigma^m(s^a)$, $a=1\ ,\dots\, p$,
    which is mapped into the physical brane $\widehat Y^\mu(s)$;
    $\widehat g$ and $\widehat A$ are the induced metric and gauge
    potential over $\widehat Y^\mu(s)$.
The integration variables in $Z$ ``live'' in the brane bulk,
while we let free the fields induced on the boundary, i.e. we do
not assign an independent classical  action to the hatted fields.

The first field to be integrated out is the gauge $p$--form $A$. The
standard routine goes through a lengthy procedure of gauge fixing
and Fadeev--Popov compensation to invert the classical kinetic
operator and define an appropriate quantum propagator.
On the other hand, one knows that
a gauge $p$--form over a $(p+1)$--dimensional manifold has no dynamical
degrees of freedom and can describe only a static interaction.
In such a limiting case the Fadeev--Popov procedure leaves no
propagating degree of freedom at the quantum level. To shorten
the whole gauge fixing procedure of ghost terms with different rank
\cite{pk}, we shall provide  an alternative ``recipe'' to kill all
the apparent degrees of freedom. We write the path--integral in the
first order version, where the gauge
potential $A$ and field strength $F$ are introduced as independent
integration variables \cite{last} and we integrate away the
gauge part of $A$ after inserting gauge fixing Dirac delta's  and the
corresponding ghost determinants in the functional measure.
The remaining, gauge invariant part of $A$ enforces $F$ to
be a classical solution of the field equations, which is a constant
background field.  No propagating degrees of freedom
survive at the quantum level. A formal proof of the
equivalence between second order and first order quantization
procedures, in the general case of a $p$--form in $p+1$ dimensions,
is beyond the purpose of this short note.  Rather,
we will briefly consider the simplest, non trivial case which is
$p=1$ gauge form over a two-dimensional, flat manifold without
boundary,
and then translate the result to  the case we are studying.
The first order, gauge fixed and Fadeev--Popov compensated path
integral is
\begin{equation}
Z_{p=1}=\int[\, DF\,] [\, DA\,]\delta\left[\, \partial^m\, A_m\,\right]
\Delta_{FP}\exp\left\{ i\int d^2\sigma \left[\, {1\over 4}F_{mn}\,
F^{mn}-{1\over 2}F^{mn}\, \partial_{[\,m} A_{n\,]}\,\right]\,\right\}
\quad .
\label{z2}
\end{equation}
By splitting $A_m$ into the sum of a ``transverse'' vector $A_m^T$
and a ``gauge part'' $\partial_m\phi$, the integration measure $[\,
DA\,]$ turns into $[\, DA^T\,][\, D\phi\,] \left(\, \det\Box
\,\right)^{1/2}$ and (\ref{z2}) reads
\begin{eqnarray}
Z_{p=1}&=&\int[\, DF\,] [\, DA^T\,] [\, D\phi\,]
\left(\, \det\Box \,\right)^{1/2}
\delta\left[\, \Box\phi\,\right]
\times
\nonumber \\
& & \qquad \qquad \times
\Delta_{FP}\exp\left\{ i\int d^2\sigma \left[\, {1\over 4}F_{mn}\,
F^{mn}-{1\over 2}F^{mn}\, \partial_{[\,m}\, A^T_{n\,]}\,\right]\,
\right\}
\nonumber\\
&=&\int [\, DF\,][\, DA^T\,]
\left(\, \det\Box \,\right)^{1/2}
\,\exp\left\{ i\int d^2\sigma\, \left[\, {1\over 4}F_{mn}\,
F^{mn}+{1\over 2}A^T_n\, \partial_m\, F^{mn} \,\right]\,\right\}
\quad ,
\end{eqnarray}
where the the gauge part has been integrated away
thanks to the Fadeev--Popov determinant $\Delta_{FP}= \det\,\Box$ and
only the gauge invariant $A^T$ vector remains in the classical action.
The extra Jacobian, coming from the change of the
integration measure, will be cancelled in a while when integrating $F$.
The pay off for relaxing the relationship between
$A$ and $F$ and getting rid of the gauge part is that  $A^T$
 {\it linearly } enters  the first
order action, i.e. $A^T$ plays the role  of Lagrange
multiplier imposing $F$ to satisfy the   classical field equations
\begin{eqnarray}
Z_{p=1}&=&\int[\, DF\,]\,\left(\, \det\Box \,\right)^{1/2}
\delta\left[\, \nabla_m\, F^{mn}\,\right]
\exp\left\{ i\int d^2\sigma \left[\, {1\over 4}F_{mn}
F^{mn}\,\right]\,\right\}\nonumber\\
&=&\int[\, DF\,]\,
\delta\left[\,  F^{mn}-\Lambda \epsilon^{mn} \,\right]
\exp\left\{ i\int d^2\sigma \left[\, {1\over 4}F_{mn}
F^{mn}\,\right]\,\right\}
\quad .
\label{z3}
\end{eqnarray}
Equation (\ref{z3}) shows that the first order formulation
of a limiting rank, abelian gauge theory, and
the Fadeev--Popov prescription lead to a  ``\textit{trivial}\,'' path
integral for $F$.
The Dirac--delta picks up  the classical configurations of
the world tensor $F$ and the whole path--integral ``collapses''
around the {\it classical trajectory\,}\footnote{The same kind of
``collapse'' around the classical trajectory has been
introduced in string theory \cite{noi} to pick up the Eguchi
``Area dynamics'' \cite{egu}. For a pedagogical introduction
to this new path--integral manipulation see
\cite{europ}, where it has been applied to a simpler
case, the non--relativistic point particle propagator.}.
Thus, $F$ is ``frozen'' to a constant value $\Lambda$ and no degrees of
freedom are left free to propagate.
The same result can be obtained, with some additional work,
for $p>1$ as well. Accordingly, we get
\begin{eqnarray}
Z &=&\exp\left\{ - {1\over p!}\int_{\partial \Sigma} dN_{k_1}\,
\sqrt{- \widehat g}\, \widehat F^{k_1\dots k_{p+1}}\,\widehat
A_{k_2\dots k_{p+1}}\,   \right\}\times\nonumber\\
& & \qquad \quad \times \int [\, DF]\,\left(\, \det\Box \,\right)^{1/2}
\delta\left[\, \partial_{m_1 }\left(\, \sqrt{-g}
F^{m_1 m_2\dots m_{p+1}}\,\right)\,\right]\times
\nonumber \\
& & \qquad \qquad \qquad \times
    \exp\left(\,{i\over 2(p+1)!}  \int_\Sigma d^{p+1}\sigma \sqrt{- g}
\, F_{m_1\dots m_{p+1}}^{\,2}\,\right)\nonumber\\
&=&\exp\left\{   -{i\Lambda\over p!}\int_{\partial\Sigma} \widehat
A_{k_1\dots k_{p+1}}\, ds^{k_1}\wedge \dots\wedge ds^{k_p}\, \right\}
\exp\left(\,-{i\Lambda^2\over 2}  \int_\Sigma d^{p+1}\sigma
\sqrt{-g} \,\right)
\quad .
\label{iord}
\end{eqnarray}
The first term in (\ref{iord}) is a pure boundary quantity produced
by a partial integration of the term $F\partial_{[\cdot}A_{\cdot]}$.
A similar term arises in string theory when boundary and bulk quantum
dynamics are properly split \cite{noi}.\\
After integrating out the gauge degrees of freedom the resulting
path--integral reads
\begin{eqnarray}
Z &=&
\int^{\widehat g} [\, Dg_{mn}\,]
\int^{\widehat Y}  [\, DY^\mu\,]
\exp\left(\, -{i\Lambda\over p!}\int_{\partial\Sigma} \widehat
A_{k_1\dots k_{p+1}}\, ds^{k_1}\wedge \dots\wedge ds^{k_p}  \,\right)
\times
\nonumber\\
&&\qquad\qquad\times\exp\left(\,-i\int_\Sigma d^{p+1}\sigma\,\left[\,
{m_{p+1}\over 2}\, {(-\gamma)\over \sqrt{-g}}
+{\Lambda^2\over 2}\sqrt{-g}\,\right]\,\right)\nonumber\\
&\equiv &\int^{\widehat g} [\, Dg_{mn}\,] \int^{\widehat Y}
[\, DY^\mu\,]\, W_{\widehat A} [\, \partial\Sigma\,]
\exp\left(\,-i\int_\Sigma d^{p+1}\sigma\,\left[\,
{m_{p+1}\over 2}\, {(-\gamma)\over \sqrt{-g}}
+{\Lambda^2\over 2}\sqrt{-g}\,\right]\,\right)
\quad .
\label{zeff}
\end{eqnarray}
We remark that this integration procedure is \textit{exact} and
leads to a bulk action plus a boundary correction represented
by the generalized Wilson factor $W_{\widehat A}[\, \partial\Sigma\,]$.
\\
We also notice that the world metric enters the path--integral only
through
the world volume density. Accordingly, we can ``change'' integration
variable
\begin{equation}
\int^{\widehat g} [\, Dg_{mn}\,]
\quad \longrightarrow \quad
\int^{\widehat g} [\, Dg_{mn}\,]\int^{\widehat e} [\, De\,]
\delta\left[\, e(\sigma) -\sqrt{-g}\,\right]
\end{equation}
and write (\ref{zeff}) as
\begin{equation}
Z=\int^{\widehat e} [\, De\,] \,\int^{\widehat Y} [\, DY^\mu\,] \,
W_{\widehat A}[\, \partial\Sigma\,]\,
\exp\left(\,-i\int_\Sigma d^{p+1}\sigma\,\left[\,
{m_{p+1}\over 2}\, {(-\gamma)\over e(\sigma)}
 +{\Lambda^2\over 2} e(\sigma)\,\right]\,\right)
\quad .
\label{zeff2}
\end{equation}
The saddle point value for the auxiliary field $e(\sigma)$ is defined
by:
\begin{equation}
 { \delta S\over \delta e(\sigma) } = 0
 \quad \longrightarrow \quad
 e_{cl.}(\sigma)={1\over\Lambda}\, \sqrt{m_{p+1}}\,\sqrt{-\gamma }
 \quad .
\end{equation}
By expanding $Z$ around the saddle point $e_{cl.}(\sigma)$
we obtain the Dirac--Nambu--Goto path--integral. Correspondingly,
we get the following {\it semi--classical} equivalence relation
\begin{eqnarray}
Z&=&\int^{\widehat g} [\, Dg_{mn}\,] \int^{\widehat Y} [\, DY^\mu\,]
\int^{\widehat A}[\, DA\,]\times\nonumber\\
&&\qquad\qquad\times\exp\left[\,  -i m_{p+1} \int_\Sigma d^{p+1}\sigma
\,{(-\gamma)\over 2\sqrt{-g}}  -{i\over 2(p+1)!}\int_\Sigma
d^{p+1}\sigma
\sqrt{- g}
\, F_{m_1\dots m_{p+1}}^{\, 2}(A)   \,\right]\nonumber\\
&\approx &
\int [\, DY^\mu\,]\, W_{\widehat A}\,
[\, \partial\Sigma\,]\exp\left[\,  - i\rho_p\,
\int_\Sigma d^{p+1}\sigma\,\sqrt{-\gamma}\,\right]
\quad .
\label{equiv}
\end{eqnarray}

The extension of the  relation (\ref{equiv}) beyond the saddle point
approximation is currently under investigation, and requires a proper
treatment of the $p$--brane degrees of freedom at the quantum level.
As is well known,
bosonic branes viewed as non--linear $\sigma$--models are
non--renormalizable perturbative quantum  field theories when $p>1$.
However, we can look at $p$--branes not as $\sigma$--models but as
elements of a more fundamental theory, say $M$-Theory, which is
in essence a non--perturbative theory. This approach is not new.
For example the
Einstein--Hilbert action in three and four dimensions
is not perturbatively renormalizable; neverthless, three
dimensional Einstein--Hilbert gravity can be reformulated as
a Chern--Simon gauge theory which can be be exactly solved at the
quantum level \cite{witten}. In a similar way, four dimensional
General Relativity can be written in terms of Ashtekar variables
which provides an exact formulation of  non-perturbative canonical
quantum gravity \cite{ash}. From the same point of view, we think that
perturbation theory is not the ultimate way to approach the problem of
brane quantization. Moreover, a
supersymmetric membrane in $D=11$ spacetime dimensions is
expected to be a finite quantum model \cite{sez}, where
both ultraviolet and infrared divergences are kept under control.
For a general discussion of quantum super-membranes we refer
to \cite{nicolai}, and  limit our considerations to the semi-classical
level. Hopefully, a proper understanding of $M$-Theory will
provide a background independent formulation of string/brane theory
where the quantum path--integral will  be well defined. In the meanwhile,
we shall work  in the WKB approximation where
one can choose the  action (\ref{lschild}), in
place of (\ref{ng}) or (\ref{spolya}), as a starting point.
The non--linearity and  reparametrization invariance of the Nambu--Goto
action  make difficult, if not impossible, to implement
the original Feynman construction of the path--integral as a sum
of {\it phase space trajectories}\cite{pol1}. One is forced, almost
unavoidably, to resort to standard perturbative approaches, e.g.
normal modes expansion, or sigma--model effective field theory.
Any perturbative approach captures some
dynamical feature and misses all the  other ones. The impressive results
obtained in string theory through duality relations of several kind
\cite{dual} show that what is not accessible in a given perturbation
scheme can be obtained through a different one. With this in mind,
we hope that an action of the type (\ref{lschild}), where the brane
variables enter polynomially, and the tension is brought in by
a generalized gauge principle, can be more appropriate to implement the
Feynman's original proposal, or, at least, to provide a {\it different
``perturbative'' quantization scheme for the Nambu--Goto model itself}.
In such a, would be, ``new regime'' of the Nambu--Goto brane both
massive and massless objects are present at once and correspond
to different values of the world manifold strength $F_{m_1... m_{p+1}}$.
It would be tempting to assign $F$ the role of ``order parameter''
and describe the dynamical generation of the brane tension as a sort
of phase transition. Furthermore, it would be interesting to extend the
action (\ref{lschild}) in order to include negative tension branes as
well.
This kind of objects appear to play an important role in the
realization of the brane world scenario \cite{rs}, \cite{nt}.\\
All these problems,  are  currently under investigation
and eventual results will be reported in future publications.

\end{document}